\documentclass[12pt]{article}

\usepackage{amsmath,amssymb}
\usepackage{amsthm}

\newcommand{\be}{\begin{equation}}
\newcommand{\ee}{\end{equation}}
\newcommand{\bea}{\begin{eqnarray}}
\newcommand{\eea}{\end{eqnarray}}

\newcommand{\gapp}{\mathrel{\raise.3ex\hbox{$>$}\mkern-14mugo
              \lower0.6ex\hbox{$\sim$}}}
\newcommand{\lapp}{\mathrel{\raise.3ex\hbox{$<$}\mkern-14mu
              \lower0.6ex\hbox{$\sim$}}}

\newcommand\lsim{\lesssim}

\newcommand\gsim{\gtrsim}

\newcommand\vev[1]{{\langle {#1} \rangle}}
\renewcommand\({\left(}
\renewcommand\){\right)}

\newcommand\del{{\mbox {\boldmath $\nabla$}}}

\newcommand\eq[1]{Eq.~(\ref{#1})}
\newcommand\eqs[2]{Eqs.~(\ref{#1}) and (\ref{#2})}
\newcommand\eqss[3]{Eqs.~(\ref{#1}), (\ref{#2}), and (\ref{#3})}

\newcommand\eqst[2]{Eqs.~(\ref{#1})--(\ref{#2})}

\newcommand\eqreff[1]{(\ref{#1})}

\newcommand\pa{\partial}

\newcommand\mpl{M_{\rm P}}

\newcommand{\dlabel}[1]{\label{#1}}

\def\calp{{\cal P}}

\def\calpz{\calp_\zeta}

\newcommand\bfk{{\mathbf k}}

\newcommand\bfn{{\mathbf n}}

\newcommand\bfp{{\mathbf p}}

\newcommand\bfx{{\mathbf x}}
\newcommand\bfy{{\mathbf y}}

\newcommand\Mpc{\,\mbox{Mpc}}
\newcommand\Gpc{\,\mbox{Gpc}}

\newcommand\sub[1]{_{\rm #1}}
\newcommand\su[1]{^{\rm #1}}

\newcommand\mone{^{-1}}
\newcommand\mtwo{^{-2}}
\newcommand\mthree{^{-3}}
\newcommand\mfour{^{-4}}
\newcommand\mfive{^{-5}}

\newcommand\half{^{1/2}}

\newcommand{\fnl}{f\sub{NL}}

\newcommand{\calpphi}{\calp_{\delta\phi}}
\newcommand{\calpphil}{\calp_{\delta\phil}}
\newcommand{\phil}{\phi\sub L}

\newcommand{\cs}{c\sub s}
\newcommand{\xls}{x\sub{ls}}

\newcommand{\kl}{k\sub L}

\begin{document}

\title{The CMB modulation  from inflation}
\author{David H.\ Lyth\\Consortium for Fundamental Physics,\\ Cosmology and
Astroparticle Group, Department of Physics,\\ Lancaster University,
Lancaster LA1 4YB, UK}
\maketitle
\begin{abstract}
Erickcek, Kamionkowski and Carroll  proposed in 2008 that the
 dipole modulation   of the CMB could be due to a very large scale
  perturbation
of the
field $\phi$ causing the primordial curvature perturbation. We repeat their calculation using  weaker
assumptions and the current data. If $\phi$ is the inflaton of {\em any} single-field
inflation with the attractor behaviour, the asymmetry
is almost certainly too small. If instead $\phi$ is {\em any} curvaton-type
field (ie.\ one with the canonical kinetic term and a negligible effect during inflation)
the asymmetry can agree with observation if
 $|\fnl|$ in the equilateral configuration is $\simeq 10$ for $k\mone=1\Gpc$ and $\lsim 3$
 for $k\mone=1\Mpc$. An $\fnl$ with these properties can apparently be obtained
 from the
  curvaton with an axionic potential. Within any specific curvaton-type
  model, the function
 $\fnl(k_1,k_2,k_3)$ required to generate the asymmetry would be determined, and
  could perhaps already be confirmed or ruled out using
 existing Planck or WMAP data.
\end{abstract}

\section{Introduction}

In 1978 Grishchuk and Zel'dovich investigated
the effect of a very large-scale enhancement of  the spectrum of the
primordial curvature perturbation, upon the CMB anisotropy \cite{gz}.
Within the observable universe the enhancement is expected to give
 an approximately linear function
of position, but the linear component has no effect upon the observed CMB anisotropy.
The leading observable effect is expected to come from the
 component that is a quadratic function of position; it  gives an enhancement to
 the CMB quadrupole, called  the Grishchuk-Zel'dovich effect, which is not
observed leading to an upper bound on the magnitude of the quadratic component.

According to present thinking the primordial density perturbation comes from
the perturbation of some field $\phi$,  which is generated from the vacuum fluctuation
during inflation.\footnote
{The perturbations of two or more fields might be involved but we do not consider that
possibility.} We are therefore talking about a very large scale contribution to the
spectrum of $\phi$.   This will generate a contribution to the spectrum of the curvature
perturbation, giving the GZ effect, but it will also generate a very large scale contribution
$\delta\phi\sub L(\bfx)$ to $\phi$. As  was
 pointed out by Erickcek, Kamionkowski and Carroll (EKC) in 2008 \cite{ekc}, the
{\em linear} component of $\delta\phil$ will
have a potentially  observable effect; it will make
 $\zeta$ statistically anisotropic within the observable universe,
  generating
a dipole modulation of the CMB anisotropy for which there is now some
  evidence \cite{wmap,planck}. I call this the EKC effect.

EKC looked  at two possibilities for $\phi$; that it is the inflaton
of slow-roll inflation       
 or that it is the  curvaton \cite{curvaton}
  with a quadratic potential.
In this paper we allow $\phi$ to be either the inflaton of
{\em any} single-field model of inflation with the attractor behaviour,
or {\em any} curvaton-type field. The latter could be the curvaton with a
generic potential    
or more generally    
any field with
 the canonical kinetic
term which has   a negligible effect during inflation.

We begin in Section \ref{s2} by recalling the dipole modulation and its
presumed origin. In Section \ref{s3} we recall the concept of a quasi-local
contribution to $\fnl$, that is generated when the potential of $\phi$
is not quadratic (self-interaction of $\phi$). In Section \ref{s4} we calculate
$A(k)$ with $\phi$ the inflaton and with $\phi$ a curvaton-type field.
 The normalization of $A(k)$ is proportional to the gradient of $\delta\phil$,
 which at this stage is not constrained.

 In Section \ref{s5} we obtain an upper bound on the gradient of $\delta\phil$,
 assuming that the observable universe occupies a typical location within a region
 that encloses all significant wavelengths of $\delta\phil$. The bound is obtained
 by requiring that (i) the GZ effect on the quadrupole is not observed and (ii)
 the expectation value of $\zeta^2$ (for a random location of the observable universe)
 is not too much bigger than 1.
 In the Conclusion we summarise our result and
 consider alternative proposals for generating the dipole modulation.
In an Appendix we describe the treatment of the GZ effect by Erickeck
et.\ al.\ \cite{ekc}, which is different from ours.

\section{CMB asymmetry from statistical inhomogeneity of $\zeta$}
\dlabel{s2}

The CMB anisotropy has been analysed to search for a dipole modulation of a
statistically isotropic quantity $\Delta T\sub{iso}$:
\be
\Delta T(\hat\bfn) = \( 1 + A \hat \bfp \cdot \hat \bfn \) \Delta T\sub{iso}
(\hat\bfn),
\dlabel{deltat} \ee
where the unit vector $\hat\bfn$ is
the direction in the sky, the unit vector $\hat\bfp$ is fixed.
`Statistically isotropic' here means that the correlators of $\Delta T\sub{iso}$ within a
disk on the sky are independent of the location of that disc.

Using \eq{deltat} for $\ell<\ell\sub{max}=64$, and $\Delta T\sub{iso}$ for higher $\ell$, Ref.\ \cite{wmap} uses WMAP data to find  $|A| =0.07\pm 0.02$.\footnote
{On a given scale we can make $A$ positive by the choice of $\hat\bfp$, but we want
to allow for a possible change in sign of $A$ going from
large to small scales, and to allow the simplest presentation of the calculation
we will not demand that $A$ is positive on large scales.}
Smoothing on a $5^\circ$ scale (corresponding to $\ell\sub{max}\sim 12$) Ref.\ \cite{planck} uses Planck data to find the same result for $A$.  Using a different method, and without making an {\em a
posteriori
} choice for $\ell\sub{max}$, Ref.\ \cite{bennet} argues that such  results are not
statistically significant, but in this paper we take them to represent a real effect.

We assume that the dipole modulation of $\Delta T$
  comes from statistical inhomogeneity of  $\zeta$ within the observable universe.
   Since $\Delta T$
depends mostly on conditions at the last scattering surface at distance $\xls=14\Gpc$
we need
\be
\zeta_\bfk(\bfx) = \( 1+ A(k) \hat\bfp \cdot \bfx/\xls +\cdots \) \zeta_\bfk(0)
,\dlabel{approx} \ee
corresponding to
\be
\calpz\half(k,\bfx) = \( 1+ A(k) \hat\bfp \cdot \bfx/\xls +\cdots\) \calpz\half(k,0)
. \dlabel{approx2} \ee
 The dots in \eqs{approx}{approx2} indicate contributions of higher order in $\bfx$,
 that must be smaller than the linear term for $x<\xls$. An equivalent definition of
 of $A(k)$ is
 \be
\frac {A(k)}{\xls} = \frac { | \del \calpz\half(k,0) | }{\calpz\half(k,0) }
. \dlabel{adef} \ee

CMB multipoles of order $\ell$ probe $k\sim \ell/\xls$
and for $k\mone$ in  the range $\xls/60$ to $\xls$, and we need $|A(k)|=0.07\pm 0.02$.
 On the much smaller scale $k\mone\sim
1\Mpc$ the distribution of distant quasars requires
$|A(k)|<0.015$ (99\% confidence
level) \cite{hirata}. Therefore, if the dipole modulation of $\Delta T$ is generated
by the dipole modulation of $\zeta$, we should write instead of
 \eq{deltat}
\be
\Delta T(\hat\bfn) = \( 1 + A(k) \hat \bfp \cdot \hat \bfn \) \Delta T\sub{iso}
(\hat\bfn),
\dlabel{deltat2} \ee
the expression applying for multipoles
  $\ell \simeq \xls k$ or equivalently angular scales
  $\Delta\theta \simeq (\xls k)\mone$.

Before continuing, we need to be precise about the meaning of \eqs{approx}{approx2}. For a cosmological perturbation $g(\bfx)$,
the correlators $\vev{g(\bfx)}$, $\vev{g(\bfx)g(\bfy)}$ and so on are
 defined as averages over some ensemble, with the actual value
  of $g(\bfx)$ corresponding to a typical realisation of the ensemble.\footnote
{We adopt the usual device of allowing $g$ to represent the actual value
or else to run over all realisations according to the context.
Correlators between different perturbations are defined in the same way.}
Under the usual assuming that the perturbation originates as a vacuum
fluctuation, $\vev{}$ is the quantum expectation value of the corresponding
operator.
If the correlators are invariant under rotations the perturbation is
said to be statistically isotropic and if they are invariant under displacements it is said to be statistically homogeneous. In the latter case the expectation values can be
taken as spatial averages for a single realisation (in particular the one corresponding
to the observed universe) so that for example $\vev{g(\bfx) g(\bfx+\bfy)}$ is the
average over $\bfy$.

The correlators should be defined within a finite box \cite{mybox}.
(This regulates long-wavelength diverges in integrals that arise when correlators
of products of perturbations are calculated. In the case of the inflationary
cosmology the box should be within the inflated patch around us.)
Within the box one uses a Fourier series that is approximated as a Fourier integral.
The region of interest should fit comfortably into the box so that physically significant
wavenumbers satisfy $kL\gg 1$ where $L$ is the box size.
For a generic perturbation
\be
g_\bfk = \int g(\bfx) e^{-i\bfk\cdot\bfx} d^3x
. \dlabel{gkdef} \ee
To define the spectrum, bispectrum etc.\ one assumes statistical homogeneity,
and we will also assume statistical isotropy.
 The
spectrum is defined by
\be
\vev{g_\bfk g_{\bfk'}} = (2\pi)^3 \delta^3(\bfk+\bfk') (2\pi^2/k^3) \calp_g(k)
, \dlabel{specdef} \ee
 the bispectrum by
\be
\vev{g_{\bfk_1}g_{\bfk_2}g_{\bfk_3}} = (2\pi^3)\delta^3(\bfk_1+\bfk_2+\bfk_3) B_g(k_1,k_2,k_3)
, \ee
and similarly for higher correlators.

In \cite{mybox}
it is proposed that the box size for cosmology should usually be the smallest one
that comfortably contains the observable universe.
 Demanding say one percent accuracy the `minimal box' size is presumably
 $L\sim 100 \xls$
corresponding to $\ln(L/\xls)\sim 5$. The use of the minimal box
avoids assumptions about
inflation long before the observable universe leaves the horizon,
and allows one to keep only the leading term when evaluating the correlators of
products of perturbations.

The minimal box is appropriate for defining the spectrum etc.\ of perturbations that can be taken to be statistically homogeneous within it.
 But to  handle
 $\zeta_\bfk(\bfx)$, one should use
 a  small box centred on $\bfx$
with size much smaller than $\xls$. Within the box, $\bfx$ can be regarded as constant and
$\zeta_\bfk(\bfx)$ can be regarded as  a statistically homogeneous
perturbation with spectrum
$\calpz(k,\bfx)$. Since the box size is $\ll\xls$,
  $\zeta_\bfk(\bfx)$ is defined only for $1/k\ll \xls$ which means that it
can only be used to describe $\Delta T$ on small angular scales corresponding to
multipoles $\gg 1$. That is as it should be, because the definition of $\Delta T\sub{iso}$ given after \eq{deltat} makes sense only on these scales.

Going to the other extreme, one can assume that the inflated patch around us is big enough to allow the use of a big box containing all significant wavelengths of
    $\zeta_\bfk(\bfx)$ considered as a function of $\bfx$ (equivalently, all significant wavelengths of
the perturbation $\delta\phil(\bfx)$) that generates $\zeta_\bfk(\bfx)$).

In this paper we first recall the generation of $\zeta$ without statistical inhomogeneity.
Then we see how to
generate the statistically inhomogeneous quantity $\zeta_\bfk(\bfx)$  using a small box. Finally, we consider
a big box which allows us to place an upper bound on the gradient of $\delta\phil$
at a typical location, taken to be our own.

\section{Generating $\zeta$ without statistical inhomogeneity}

\dlabel{s3}

In this section we recall the standard description of $\zeta$ and its generation
from the perturbation of some field $\phi$.
 We begin with the definition of $\zeta$, which makes no reference
to its stochastic properties (and therefore invokes no box).
The curvature perturbation $\zeta$ is taken to be smooth on some comoving scale
$x\sub{smooth}$ that is shorter
than any of interest (ie.\ it is taken to only have modes with
$k\lsim x\mone\sub{smooth})$.\footnote
{To be more precise its gradient is $\lsim x\mone\sub{smooth}$.
We are not really invoking
a Fourier transform here but use $k$ for ease of presentation. The same device will be
used later without comment.}
Also, $\zeta$
 is defined only while the smoothing  scale is outside the horizon
($x\sub{smooth}\gg 1/aH$).
Using the comoving threads of spacetime and the slices of constant energy density,
$\zeta$ is defined by
\be
\zeta(\bfx,t) \equiv \delta ( \ln a(\bfx,t) ) \equiv \ln[a(\bfx,t)]-\ln[a(t)]
, \dlabel{zetadef} \ee
where $a(\bfx,t)$ is the scale factor such that a comoving volume element
has volume $\propto a^3$. Here $a(t)$ is the scale factor in the background universe
that is invoked to define perturbations.

In the early universe $\zeta$ may be time-dependent,
but if we take the smoothing scale to be the shortest cosmological scale it
has reached some  time-independent value $\zeta(\bfx)$
 at least by the time that the smoothing scale is approaching horizon
entry.
(By `cosmological scales' we mean those that are probed by the CMB anisotropy
and high redshift galaxy surveys, corresponding $e^{-15}\xls <k\mone<\xls$.)
If $\phi$ is the inflaton of single-field inflation, $\zeta$
already has the final value soon after $x\sub{smooth}$ leaves the horizon;
 in the opposite case that $\phi$ is a curvaton-type
field the final value  is reached only at some epoch after inflation.

The  curvature perturbation $\zeta(\bfx,t)$
generated by $\phi$ is given by the non-linear $\delta N$  formula \cite{ss,lr}
\be
\zeta(\bfx) \equiv \delta (\ln a(\bfx,t) )= \delta (\ln a(\bfx,t)/a(t_1) )
\equiv \delta N(\phi(\bfx,t_1))
. \dlabel{deln1} \ee
The function $N(\phi(\bfx,t_1))$
is the number of $e$-folds of expansion at position $\bfx$
between a time $t_1$ during inflation after the smoothing scale has left the horizon,
with $\phi$
has the  assigned value, and  a time $t$ at which the energy density has a
fixed value. The field $\phi(\bfx,t_1)$ is defined on a `flat' slice of spacetime
(one on which the scale factor has a fixed value), and $\zeta$ is independent of
the choice of $t_1$.

The field is written
\be
\phi(\bfx,t) =  \phi_0(t) + \delta\phi(\bfx,t)
. \ee
Expanding \eq{deln1} gives \cite{lr}
\be
\zeta(\bfx) = N'(\phi_0(t_1)) \delta\phi(\bfx,t_1)
+ \frac12 N''(\phi_0(t_1)) (\delta\phi(\bfx,t_1))^2 +\cdots
\dlabel{zetaofphi} .\ee

The spectrum $\calpz$ and bispectrum $B_\zeta$ are  defined by \eqs{specdef}{bdef}. Instead of the latter one usually works with
\be
\fnl(k_1,k_2,k_3)\equiv \frac56\frac{B_\zeta(k_1,k_2,k_3)}
{P_\zeta(k_1)P_\zeta(k_2) + 2\mbox{\,perms} }
, \dlabel{fnldef} \ee
where $P_\zeta(k)\equiv (2\pi^2/k^3)\calpz(k)$.

Analyses of the data to obtain observational constraints on $\calpz$ and $\fnl$  take $\zeta$ to be statistically homogeneous and
isotropic within a  box of at least minimal size.
The constraints are obtained on the assumption that our location within
the chosen box is typical, and they take $\Delta T$ to be statistically isotropic.
For $\calpz(k)$,
 observation \cite{wmap2,planck2} gives $\calpz\half\simeq5\times 10\mfive$ and
\be
\frac{n(k)-1}2 \equiv \frac1{\calpz\half(k)} \frac {d\calp\half(k)}{d\ln k}
=-0.040\pm 0.007
. \ee
If $\fnl$ is independent of $k_i$ then  \cite{planck3}
 $\fnl=2.7\pm 5.8$,
 but the bound is
much weaker for generic $k_i$, roughly  $|\fnl|\lsim 100$.
If $|\fnl|\gsim 1$ it will eventually be detected.

Except where
stated  we assume that $\phi$ has the
canonical kinetic term  when these  scales are leaving
the horizon during inflation.
Then  $\delta\phi_\bfk$
is created from the vacuum fluctuation
at the epoch of horizon exit  $aH=k$.
The spectrum is initially
  $\calpphi(k)=(H/2\pi)^2$ and
$B_{\delta\phi}(k,k,k)$ is initially very small.

To keep $\fnl$ within the observational bound,
 the first term of \eq{zetaofphi} must
dominate giving
\be
\calpz(k) = N'^2(\phi_0(t_1)) (H(t_1)/2\pi)^2
.\ee
Including the second term,
$\fnl$ is given by
\cite{lr,bntw}
\be
\frac65 \fnl(k_1,k_2,k_3) = \frac{N''(t_1)}{N'^2(t_1)}
+ \frac{(2\pi)^3 N'^3(t_1)B_{\delta\phi}(k_1,k_2,k_3,t_1)}
{P_\zeta(k_1)P_\zeta(k_2) + 2\,\mbox{perms} }
\dlabel{fnlexpression} . \ee

For our purpose we can set $t_1=t_k$ where $t_k$ is
 horizon exit  for a scale $k$ which is taken to be the smoothing scale.
Then
\be
\calpz(k) = N'^2(\phi_0(t_k)) (H(t_k)/2\pi)^2
\dlabel{pzoftk} ,\ee
and
\be
\fnl\su{qlocal}(k) = \frac 56 N''(\phi_0(t_k))/N'^2(\phi_0(t_k))
, \dlabel{fnllocal} \ee
where $\fnl\su{qlocal}(k)\equiv\fnl\su{qlocal}(k,k,k)$  and
the superscript qlocal (quasi-local) means that
$\delta\phi$ is taken to be  gaussian at horizon exit so that
$B_{\delta\phi}(k,k,k,t_k)=0$. (If $\fnl\su{qlocal}$ is independent of $k$
it is called the
local contribution.)

We suppose  first  that $\phi$ is the inflaton of slow-roll inflation \cite{sr}.
In this case, $\zeta_\bfk(t)$ achieves its final value promptly at $t=t_k$,
which means that
\be
\zeta_\bfk= H(t_k) (\delta t)_\bfk =- H(t_k) \delta\phi_\bfk(t_k)/\dot\phi_0(t_k)
, \dlabel{35}\ee
where $\delta t$ is the displacement of the slice of uniform energy density from the
flat slice on which $\delta\phi$ is defined, and the second equality is valid to first
order in $\delta\phi_\bfk$.  This gives
\bea
\calpz(k) &=& \( \frac {H(t_k)} {\dot\phi_0(t_k)} \)^2 \calpphi(k,t_k)
\dlabel{36} \\
\calpphi(k,t_k) &=& \( \frac{ H(t_k)}{2\pi} \)^2\dlabel{37}
. \eea
The slow-roll approximation corresponds to conditions on the scalar
field potential,
$\epsilon\ll 1$ and $|\eta|\ll 1$
 where    $\epsilon\equiv \mpl^2 (V'/V)^2/2$
and $\eta\equiv \mpl^2 V''/V$,
and
\be
 \dot\phi_0(t_k) \simeq - V'(\phi_0(t_k)/3H(t_k) \dlabel{39}
.\ee
These imply
\be
3\mpl^2 H^2(t_k)  \simeq V(\phi_0(t_k)) \dlabel{38}.
\ee

These equations make $\calpz(k)$ a function of $\phi_0(t_k)$.
Using $k=a(t_k) H(t_k)$ and the good approximation $d\ln k\simeq  d\ln a = Hdt$
this gives
\bea
\frac{n(k)-1}2 &=&
\frac1{\calpz\half(k)}
 \frac {d\calpz\half(k)}{d\phi} \frac{\dot\phi}H
\dlabel{nmoneq} \\
&\simeq& \eta(t_k) - 3\epsilon(t_k)
    \dlabel{nofetaepsilon} .\eea
Evaluating \eq{fnllocal} gives
\be
    \frac65\fnl\su{qlocal}(k) \simeq   2\epsilon(t_k) - \eta(t_k)
    . \ee
    Barring a
fine-tuned cancellation
  \be
|\fnl\su{qlocal}(k)|  \sim 1-n(k)
  \dlabel{fnllocalbound} \ee
  and in any case $|\fnl\su{qlocal}|\ll 1$.
These expressions assume $B_{\delta\phi}(k,k,k,t_k)=0$, which is not a good approximation
with $\phi$ the inflaton because the first term of \eq{fnlexpression} is also very small.
 The full $\fnl$  \cite{srng} still satisfies
 $|\fnl|\ll 1$
but we don't need it.

Instead of slow-roll inflation one can consider
the most general single-field inflation paradigm, in which the unperturbed solution
$\phi(t)$ is unique up to a time translation; in other words, $\dot\phi$ is a function
of $\phi$ (attractor behaviour). Within this paradigm, one can still expect \eqs{nmoneq}{fnllocalbound}
to apply. We have verified this explicitly for the case of k-inflation
\cite{kinflation}.
The Lagrangian is an arbitrary function of $\phi$ and $\pa_\mu \phi \pa^\mu \phi$,
and we will choose $\phi$ so that $\delta\phi$ has the canonical action for a free
 field.\footnote
{In the notation of \cite{kinflation}, $\delta\phi=v$.}
The epoch $t_k$ at which $\delta\phi$ is generated from the vacuum fluctuation
is given by $\cs k=aH$, and
\bea
\calp_{\delta\phi}(t_k,k) &=& \cs\mthree (H(t_k)/2\pi)^3 \\
\calpz(k) &=& \frac1{2\cs}\frac{H^2}{\mpl^2 |\dot H|} \( \frac H{2\pi} \)^2
.\eea
{}From these and \eq{35}  we deduce for $N_\phi\equiv dN/d\phi$
\be
N^2_\phi(t_k) = \frac{\cs^2(t_k) H^2(t_k)}{2\mpl^2 \dot H(t_k)} =
\frac{H^2(t_k)}{\dot\phi^2(t_k)}
. \ee
The conditions $|\dot H/H^2| \ll 1$, $|\ddot H/\dot H H|\ll 1$
and $|\dot\cs/H\cs|$ are imposed and we then have
\bea
n(k)-1 &=& - \frac{\dot \cs}{\cs H} + 4\frac{\dot H}{H^2} - \frac{\ddot H}{\dot H H} \\
\frac 65 \fnl\su{qlocal}(k) &=&
\frac{\dot \cs}{\cs H} + \frac{\dot H}{H^2} - \frac12\frac{\ddot H}{\dot H H}
. \eea
(The final relation has not been given before.)

In summary, \eq{nmoneq} is satisfied for k-inflation, and so is
\eq{fnllocalbound}  barring a cancellation, and  in any case
$|\fnl\su{local}(k)|\ll 1$. The first equation holds whenever the epoch $t_k$
at which $\delta\phi $ is generated satisfies $k=f(t_k) a(t_k) H(t_k)$
with $|\dot f/f| \ll H$. The second equation holds if $n(k)-1$ and $\fnl\su{local}(k)$
are both linear combinations of small parameters with numerical coefficients roughly
of order 1, and $|\fnl\su{local}(k)|\ll 1$ then in any case holds.
One can expect all of these features for  any single-field inflation model with the attractor behaviour.

Finally, consider  the case that $\phi$ is a curvaton-type field.
In this case  one expects barring cancellations
$|\fnl\su{qlocal}(k)|\gsim 1$, which we will
assume. (As we will see, $|\fnl\su{qlocal}(k)|\sim 10$ is actually required to generate the required asymmetry for $k\mone \sim \Gpc$.)
Then $\fnl\su{qlocal}$ is practically equal  to the full quantity $\fnl$
\cite{ourng} and we will identify them. Also, the  evolution of $\delta\phi_\bfk$
 after horizon exit is given by
  \be
H(t)  \dot\delta\phi_\bfk(t) =
 - V''(\phi_0(t))  \delta \phi_\bfk(t)- \frac12 V'''(\phi_0(t))
[(\delta \phi(t))^2]_\bfk - \cdots
. \dlabel{curvatonevolution} \ee
If $V$ is quadratic,
 the evolution is linear. Then  $B_{\delta\phi}(k,k,k)$ remains
 zero if it is zero initially, which from \eq{fnlexpression} means that
$\fnl(k)$ is a constant.

\section{Generating  $\zeta_\bfk(\bfx)$}

\dlabel{s4}

Now we write
\bea
\phi(\bfx,t) &=&  \phi_0(\bfx,t) +  \delta\phi(\bfx,t) \\
\phi_0(\bfx,t) &=& \phi_0(t) + \delta\phil(\bfx,t)
, \eea
where $\delta\phi$ has $k>\xls\mone$ and $\delta\phil$ has
$k< \xls\mone$. For the curvature perturbation we write
\bea
\zeta(\bfx) &=& \zeta_{\delta\phi} (\bfx) + \zeta\sub{GZ}(\bfx) \dlabel{zsgz} \\
\zeta_{\delta\phi}(\bfx) &\equiv&  N(\phi(\bfx,t_k)) - N(\phi_0(\bfx,t_k))\nonumber \\
&=& N'(\phi_0(\bfx,t_k))\delta\phi(\bfx,t_k) + \cdots \dlabel{15}  \\
\zeta\sub{GZ}(\bfx) &\equiv&   N(\phi_0(\bfx,t_k) ) - N(\phi_0(t_k)) \nonumber \\
&= & N'(\phi_0(t_k)) \delta\phil(\bfx,t_k) +
\frac12N''(\phi_0(t_k))\( \delta\phil(\bfx,t_k) \)^2+\cdots  \dlabel{17}
. \eea

Each of the contributions $\zeta_{\delta\phi}$ and $\zeta\sub {GZ}$ is independent of $t_k$
because they vary on different scales and $\zeta$ itself is independent of $t_k$.
The first term of \eq{15} dominates because $\zeta_{\delta\phi}$ is almost gaussian,
and we  assume  that the  first term of \eq{17} dominates which will be justified
in the next section. Then
$\delta\phil(\bfx,t_k)\propto 1/N'(\phi_0(t_k))$, which from \eq{pzoftk}
is proportional to $H(t_k)$.
Analogously with \eqs{approx2}{adef}  we write
\bea
\delta\phil(\bfx,t_k) &=& B(k)  (H(t_k)/2\pi) \hat\bfp\cdot\bfx/\xls +\cdots\dlabel{bdef1} \\
(H(t_k)/2\pi)  B(k) /\xls &\equiv& |\del (\delta\phil(\bfx,t_k))|\sub{x=0}
\dlabel{bdef} . \eea
(Remember that $t_k$ is a function of $k$ so that either can be used as an argument.)

In the next section, we derive an upper bound on $B$ on the assumption that the observable universe occupies a typical position, within a region big enough to contain all significant wavelengths of $\delta\phil$. If $\phi$ is a curvaton-type field the bound is
\be
\calpz(k) B^2 \lsim  4\times 10\mfour/|\fnl(k)|
\dlabel{ekcbound}. \ee
If instead $\phi$ is the inflaton, the right hand side is just $4\times 10\mfour$.

Both $\zeta_{\delta\phi}$ and $\zeta\sub{GZ}$ contribute to the CMB quadrupole,
but the GZ effect that might have enhanced the quadrupole comes only
from $\zeta\sub{GZ}$. We deal with it in the next section, but for now focus
on $\zeta_{\delta\phi}$.
Evaluated  in a  box with size $\ll \xls$ centred at position $\bfx$ it gives
 \bea
 \zeta_\bfk(\bfx) &=& N'(\phi_0(\bfx,t_k)) \delta\phi_\bfk(\bfx,t_k) + \cdots \\
 \calpz(k,\bfx) &=&  N'^2(\phi_0(\bfx,t_k)) ((H(\bfx,t_k)/2\pi)^2
 \dlabel{zkofx}  ,\eea
 where $H(\bfx,t)\equiv \dot a(\bfx,t)/a(\bfx,t)$ is the unperturbed quantity
 within the small box and we kept only the first term in evaluating \eq{zkofx}.
 After insertion into \eq{adef}, this gives $A(k)$.
 We will take the results of the previous section to apply to $\zeta_\bfk(0)$
 and $\calpz(k,0)$.

We consider first the case that $\phi$ is the inflaton of single-field  inflation.
Then $\calpz(k,\bfx)$ is a function of $\phi_0(\bfx,t_k)$ and
using \eq{nmoneq} we have
\be
\frac{\del \calpz\half(k,0)}{\calpz\half(k,0)} = \frac{(1-n(k))}2
\frac {H(t_k)}{\dot\phi_0(t_k)} \del \phi(0,t_k)
\dlabel{41a}.\ee
Using \eqs{adef}{bdef} this gives
\be
A(k) = \frac{1-n(k)}2 B \calpz\half(k)
\dlabel{aforinf}. \ee
Observational constraints on $n(k)$ easily allow $A(k)$ to have sufficient scale
dependence \cite{wmap2,planck2}, but the bound $\calpz(k) B^2 \lsim 2\times 10\mfour$
makes $A(k)$ too small.

Now  suppose instead that $\phi$ is a curvaton-type field.
Since
$\phi$ has a negligible effect during inflation, $H(\bfx,t_k)$ is independent
of $\bfx$ and \eqs{adef}{bdef}
\eq{zkofx} give
\be
A(k)=
 \frac65 f\sub{NL}(k) B \calpz\half(k)
. \dlabel{acurv} \ee

Using \eq{ekcbound},
\be
|A(k)| \lsim 0.018 |\fnl(k)|\half
. \ee
To have
 $|A(k)|= 0.07\pm 0.02$
on the $\Gpc$ scale we need $|\fnl(k)|\gsim 8$ on that scale.
A tight observational bound on $\fnl(k)$ could be obtained using
a shape for $\fnl(k_1,k_2,k_3)$ derived within a specific curvaton-type
model (see \cite{bntw,axioncurv} for the curvaton)
but it would presumably be no  tighter than the
result $|\fnl|\sim 10$ that holds if $\fnl$ is a constant.
We conclude that
 the linear GZ effect can
account for the CMB asymmetry if $\zeta$ is generated by
 a curvaton-type field.

Before leaving this section we mention a perhaps simpler way of proceeding when
$\phi $ is the curvaton. Instead of
 \eqst{zsgz}{17} one can write
\bea
\zeta(\bfx) &=&  N(\phi(\bfx,t_k) - N(\phi_0) \\
&=& N'(\phi_0(t_k)) \( \delta \phi(\bfx) + \delta\phil(\bfx) \)
+ \frac12 N''(\phi_0(t_k)) \( \delta \phi(\bfx) + \delta\phil(\bfx) \)^2
+\cdots \\
&\equiv&  \( \zeta\sub S(\bfx) + \zeta\sub L(\bfx) \) + \frac35 \fnl(k)
\( \zeta\sub S(\bfx) + \zeta\sub L(\bfx) \)^2 + \cdots
\dlabel{alternative1} \\
&=&\( 1 + \frac65 \fnl(k) \zeta\sub L(\bfx) \) \zeta\sub S(\bfx)
+ \zeta\sub L(\bfx) + \frac35 \fnl(k) \zeta^2\sub L(\bfx) +\cdots
, \dlabel{alternative} \eea
where
$\zeta\sub S \equiv  N'(\phi_0(t_k)) \delta\phi(\bfx)$
and $\zeta\sub L \equiv N'(\phi_0(t_k)) \delta \phil(\bfx)$.
The first term of \eq{alternative} corresponds to $\zeta_{\delta\phi}$
of \eq{zsgz} and the other two terms correspond to $\zeta\sub{GZ}$ of
\eq{zsgz}.
With $\phi$ the inflaton, \eq{alternative} is still correct, but
not very useful for calculating $\calpz(k,\bfx)$ because
$\calpphi(k,t_k)$ within a small box depends on the position $\bfx$.

\section{The view from a big  box}
 \dlabel{s5}

The calculation of the previous section we invoked only the observable universe corresponding to $x<\xls$. The function $\delta\phil(\bfx)$ was taken as a given quantity
without discussing its origin.

In this section we assume that the nearly homogeneous patch containing the
observable universe contains all significant wavelengths numbers of $\delta\phil(\bfx)$. That is desirable because it allows $\delta\phil$
to be generated from the vacuum fluctuation like
 $\delta\phi$. With this assumption we will derive \eq{ekcbound} if $\phi$ is a curvaton-type field, and the same bound without the $\fnl$ factor if $\phi$ is the inflaton.

 Within the big box, $\zeta$ is statistically homogeneous.
To proceed, we use \eq{alternative1},
choosing  $t_k=t_1$ where $t_1$ is the epoch of horizon exit for the scale
$k_1\mone =1\Gpc$.

 We assume that the first term of \eq{alternative1} dominates.
This makes  $\calpz(k)=N'^2 \calp_{\delta\phi}(k)\simeq (5\times 10\mfive)^2$ for
$k>\xls\mone$,
and $\calpz(k)=N'^2 \calp_{\delta\phil}(k)$ for $k<\xls$.
Also, since $\zeta_\bfk \simeq N'\delta\phi_\bfk$ on cosmological scales,
it is at least approximately gaussian on those scales, though its
 bispectrum etc.\ within the big box cannot be calculated without further
 assumptions (ie.\ we do not know how good is the approximation
 $\zeta_\bfk \simeq N'\delta\phi_\bfk$).

If $\phi$ is a curvaton-type field we expect  $|\fnl\su{local}(k_1)|
\simeq |\fnl(k_1)| \gsim 1$. Then the
 condition that the first term of \eq{alternative1} dominates
 corresponds to
$\fnl^2(k_1) \vev{\zeta^2}\lsim 1$.
This is a bit stronger than the condition
$\vev{\zeta^2}\lsim 1$ that is usually imposed when discussing the GZ effect.
If instead $\phi$ is the inflaton,
$\
|\fnl\su{qlocal}(k_1)|\ll 1$ and  it is weaker.

The condition $\vev{\zeta^2}\lsim 1$
is not strictly required because a nearly constant value of $\zeta$
in the observable universe can be absorbed into the scale factor $a(t)$. However,
we do require $\calpz(k)\lsim 1$
so that the spatial curvature scalar within a region with size $k\mone$
is $\lsim k$. Indeed, a violation of that condition would imply a strong spatial curvature
which would invalidate the interpretation of $\bfx$ as a distance, for a typical
region which we are supposed to occupy \cite{kopp}.

Dropping the small short-scale contribution   we have
\be
\vev{\zeta^2} = \int^{\xls\mone}_0 \frac {dk}k \calpz(k)
. \ee
We see that $\calpz(k)\lsim 1$ implies at least roughly
 $\vev{\zeta^2}\lsim 1$ unless
the integral receives significant contributions from a very large range
 $\Delta \ln k \gg 1$. We will see that this would probably make $A(k_1)$ too small.
 We therefore assume
\be
 \int^{\xls\mone}_0 \frac {dk}k \calpz(k) \lsim  \fnl\mtwo(k_1)
,  \dlabel{fnlcon} \ee
if $\phi$ is a curvaton-like field with $\fnl^2(k_1)\gsim 1$.
If instead $\phi$ is the inflaton we set the right hand side equal to 1.
In both cases the first term of \eq{alternative} dominates for a typical value of $\delta\phil(\bfx)$, which means that the first term of \eq{17} dominates as advertised.

We assume that our location within the big box is typical.
Multiplying both sides of
 \eq{bdef} by $N'$ and  squaring them gives
\be
\calpz(k_1) B^2 \simeq  \int^{\xls\mone} _0 \frac{dk}k (\xls k)^2 \calpz(k)
.\dlabel{typical}\ee

 For the CMB multipoles, $a_{\ell m}^2 \simeq \vev{a_{\ell m}^2}
 \equiv C_\ell$ with
\be
C_\ell=4\pi \int_0^\infty T_\ell^2(k) \calpz(k) dk/k
, \dlabel{cell} \ee
where  $T_\ell(k)$ is $\sim 1$ for $k\mone \sim \xls$
and close to 1 for $k\mone\gg \xls$.

The cosmic variance of $a^2_{\ell m}$ is defined as the mean-square difference
between $a^2_{\ell m}$ and $C_\ell$ (ie.\ as $\vev{ (a_{\ell m}^2
-C_\ell)^2 }$). If $\zeta_\bfk$ were  gaussian, $a_{\ell m}$
would have a gaussian probability distribution and the cosmic variance
would be $2C_\ell^2$. Since $\zeta_\bfk$ is at least approximately gaussian
we expect that to be at least approximately correct, but the precise cosmic variance cannot be calculated
without further information.
The observed dipole modulation corresponds to a systematic bias of the observed $a_{\ell m}^2$
away from $C_\ell$, and \eq{typical} will ensure that
the bias is allowed (for a typical observer) by the cosmic variance of $C_\ell$.
An investigation of how that  comes about
is beyond
 the scope of this paper.

Using the Sachs-Wolfe approximation, the GZ contribution to $C_2$ is
\be
C_2\su{GZ} = \frac{4\pi}{25} \int_0^{{\xls\mone}}  dk \( \frac{(k \xls)^2 }{15} \)^2 \calpz(k)
. \ee
    We  will require $\sqrt{C_2\su{GZ}}$ to be $\lsim$  3 times the rms quadrupole found in \cite{george}, giving
\be
 \int_0^{{\xls\mone}}  \frac{dk}k (k \xls)^4 \calpz(k)
\lsim   (3.8 \times 10\mfour)^2
\dlabel{c2gz} . \ee

{}{}Using the Cauchy-Schwartz inequality, \eqss{fnlcon}{typical}{c2gz} imply
if $\phi$ is the curvaton
\be
B^2\calpz(k_1) \lsim 4\times 10\mfour |\fnl(k_1)|\mone
. \dlabel{thatbound} \ee
The result for $\phi$ the inflaton is obtained by setting $\fnl(k_1)=1$.

The bound \eqreff{thatbound}  is saturated by choosing $\calpz(k)\propto \delta(k-\kl)$ with
$(\xls\kl)^2\simeq 4\times 10\mfour |\fnl(k_1)|$.
Reducing   $\xls\kl$ by a factor of 10
  gives $B^2\calpz(k_1) \lsim 10^{-6}$ which makes $A(k_1)$ much too small.
  The same is true if we increase it by that factor, though this might be regarded as
  incompatible with $\xls\kl \ll 1$.

Instead of a strong peaking one might assume a flat plateau:
$\calpz(k)$ constant in a range $\ln \kl - N\sub L < \ln k < \ln\kl$
with $N\sub L\gg 1$. Then the weakest bound is  for $(\xls\kl)^2
\simeq 8\times 10\mfour |\fnl(k_1)| \sqrt{N\sub L}$ which gives
\be
2N\sub L B^2 \calpz(k_1) |\fnl(k_1)| \lsim 4\times 10\mfour
.\ee
Since $|\fnl(k_1)|\lsim 100$ we need $N\sub L\lsim 5$.
As  before a value of
$\xls\kl$ much below $10\mtwo$ is not allowed.

These examples suggest that we need $\calpz(k)$
to grow sharply below some  $k=\kl\sim 10\mtwo \xls\mone$, and to fall off
when $k\mone $ is not far below $\kl$.
It remains to be seen if this allows a plausible mechanism for generating
the required large $\delta\phil$  from the vacuum fluctuation.

We  close this section by mentioning a different possibility for obtaining a spatial variation
of the background field $\phi_0(\bfx)$. Instead of invoking an enhancement of $\calpphil(k)$, we can keep the usual fairly flat spectrum and suppose
that we live at a special place within the  large box. This possibility was
considered in detail by Linde and Mukhanov \cite{lm} for the curvaton with
a quadratic potential, who noticed that it could
generate dipole modulation of the CMB. It is not clear how easily this
scheme could keep the CMB quadrupole small enough, while generating the required
asymmetry.

\section{Conclusion}
\dlabel{s6}
It was disappointing, though hardly a  surprise, that the
GZ effect  was not seen when the  CMB quadrupole was first observed.
The dipole modulation of the CMB anisotropy may now be
making up for that disappointment, by exhibiting the closely related
 EKC effect.

On the assumption that $\zeta$ is generated by some field $\phi$, we have found that
the EKC effect almost certainly
cannot  generate the observed asymmetry if $\phi$ is the inflaton,
but that it can do so if $\phi$ is a curvaton-type field.
This is perhaps the first
indication that the latter may be nature's choice.
The  asymmetry can agree with observation if
 $|\fnl|$ in the equilateral configuration is $\simeq 10$ for $k\mone=1\Gpc$ and $\lsim 3$
 for $k\mone=1\Mpc$. An $\fnl$ with these properties can apparently be obtained
 from the
  curvaton with an axionic potential \cite{axioncurv}. Within any specific curvaton-type
  model, the function
 $\fnl(k_1,k_2,k_3)$ required to generate the asymmetry would be determined, and
  could perhaps already be confirmed or ruled out using
 existing Planck or WMAP data.

 It remains to be seen if a plausible
 mechanism can be found for generating $\delta\phil$ from the vacuum
 fluctuation. On the other hand, it seems hard to come up with
an alternative to the EKC effect.
 After ruling out various scenarios, \cite{pesky} mention only five  that might still
 be viable.\footnote
 {They rule out the large-scale non-gaussianity proposal of \cite{schmidt} on the ground
 that it makes $A(k)$ scale-independent, but this scenario in fact generates only
 statistical anisotropy of $\calpz$ which cannot generate the CMB  asymmetry.
  Figure 1 of that paper refers to $L=2$ in their notation, not to $L=1$
 as stated in the caption. I thank Fabian Schmidt for clarification of this issue.}
These are (i) an inhomogeneous tilt  $n(k,\bfx)-1$, (ii) a
statistically inhomogeneous  isocurvature perturbation  \cite{iso},
(iii) a statistically inhomogeneous  tensor perturbation, (iv)  asymmetry of the
optical depth, (v) bubble collisions and (vi) non-trivial topology of the Universe.
Of these, the first is identical with our version of the EKC effect (with possible generalisation  to the case that $\zeta$ comes from a curvaton-type field with
a non-canonical kinetic term, or from
two or more field perturbations) and the second
may be in conflict with Planck bounds on the isocurvature amplitude. The next two are
only partially investigated in \cite{pesky} and may also be in conflict with existing data
while the last two have not been tried at all.
There is also the  proposal of \cite{cw}, which replaces, during inflation, the usual
 Riemannian spacetime by what is called Randers spacetime. The
 $A(k)$ appears to be
viable, with $A\propto 1/k$.\footnote
{This dependence will be explained in a future version of \cite{cw}.
(Personal communication from S. Wang.)}

Since the first version of this paper appeared on arXiv.org there have been four more papers.
That of \cite{lgp}  invokes a contraction of the universe, followed by
an inflationary expansion with $\phi$ the inflaton.
On scales leaving the horizon during the contraction, $\dot H/H^2$ is enhanced which sufficiently enhances
 $A(k)$. But the enhancement applies only to scales
 $k\mone > ({\cal H}_0)\mone$ (their notation) which for their best fit corresponds to
 $k\mone> 5.4\Gpc\sim \xls/3$. That is outside the
required range  $\xls/60 \lsim k\mone \ll \xls$.
(At   $3\sigma$ though, they can have $1/{\cal H}_0 = \xls/30$ which might work.\footnote
{Personal communication from Yun-Song Piao.})
In \cite{john} an implementation of the isocurvature scenario of \cite{iso} is
proposed.
In \cite{anupam,hassan} they consider the EKC effect with, among
other things, the possibility of
two or more curvaton-type fields.

Although all possibilities should be explored, it
 seems fair to say  that there is at present no
proposal which looks more plausible than the EKC effect.

 \section*{Acknowledgements}
 I thank Andrew Liddle for valuable comments.
 The work  is supported by the Lancaster-Manchester-Sheffield Consortium for
  Fundamental Physics under STFC grant ST/J000418/1.

\appendix
\section{The EKC treatment of the GZ effect}

In \cite{ekc} the GZ effect is treated in a way that is  different and less general
than ours. For the case that $\phi$ is the inflaton of slow-roll inflation, they consider only the first term
of $\zeta\sub{GZ}$ because they work to first order in $\phi$. In that term they
take $\delta\phil$ to be sinusoidal, with the $B=0$ so that the leading GZ effect is
for the octupole. Requiring it to be less than the observed quantity they conclude that
$A(k)$ is too small.

For the case that $\phi$ is the curvaton with a quadratic potential, they
include also the second  term of $\zeta\sub{GZ}$.\footnote
{I thank A.\ Erickcek for pointing this out to me.}
Inserting \eq{bdef1} this gives
 \bea
 \zeta\su{EKC}\sub{GZ}(\bfx,t_k)
 &=& \frac12 N''(\phi_0(t_k))B^2
(H(t_k)/2\pi)^2 \xls\mtwo (\bfx\cdot\hat p)^2 \\
 &=& \frac35 |\fnl(k)| B^2 \calpz(k) \xls\mtwo (\bfx\cdot\hat p)^2
.\eea
Using Eqs~(2)--(4) of \cite{ekc}  with $\Phi=-(3/5) \zeta$, this gives
a contribution to
 the quadrupole  given by
\be
| \fnl(k)| B^2 \calpz(k) = 2.2\times 10\mfour
 \(\frac {|a_{20}\su{EKC}|}{1.8\times 10\mfive} \)
 ,\dlabel{aekc}\ee
 where the polar axis for $a_{20}$ is along the $\hat \bfp$ direction.
 Barring a cancellation, $|a_{20}\su{EKC}|$ should be
 $\lsim$  the observed $|a_{20}|$.
  That has yet to be extracted from the data,
 and EKC required instead that $|a_{20}\su{EKC}|$ be less than 3 times the rms
 value of $a_{\ell m}$ found in \cite{george}
 corresponding to $|a_{20}\su{EKC}|<1.8\times 10\mfive$.
With that assumption, \eq{aekc} gives
 \be
|\fnl(k)| B^2 \calpz(k) < 2.2\times 10\mfour
. \ee
This is  essentially the same as our  bound \eqreff{thatbound},\footnote
{The difference probably comes from our use of the Sachs-Wolfe approximation
as opposed to their exact evaluation of the quadrupole.}
but its status is very different because it ignores the first term of $\zeta\sub{GZ}$.
As we have seen, the total GZ contribution to the quadrupole can be much smaller
than $a_{20}\su{EKC}$, indicating a cancellation between the first and second
terms of $\zeta\sub{GZ}$. In that regime, our bound  \eqreff{c2gz} is
a consequence of $\fnl(k_1)\vev{\zeta^2}\lsim 1$ which has nothing to do with the
GZ effect.

\end{document}